\pdfoutput=1
\documentclass[11pt]{article}
\usepackage{acl}
\usepackage{graphicx}
\usepackage{times}
\usepackage{latexsym}
\usepackage{amsmath}
\usepackage{amssymb}
\usepackage[T1]{fontenc}

\usepackage{placeins}
\usepackage{booktabs}

\usepackage[utf8]{inputenc}

\usepackage{microtype}

\usepackage{inconsolata}
\usepackage{marvosym}
\usepackage{graphicx}

\title{Beyond ESM2: Graph-Enhanced Protein Sequence Modeling with Efficient Clustering}

\author{
Shujian Jiao$^1$~~
Bingxuan Li$^2$~~
Lei Wang$^1$~~
\\
\textbf{Xiaojin Zhang}$^1$~~
\textbf{Wei Chen}$^1$\textsuperscript{\Letter} ~~
\textbf{Jiajie Peng}$^3$~~
\textbf{Zhongyu Wei}$^1$\textsuperscript{\Letter} ~~
\smallskip
\\
$^1$Huazhong University of Science and Technology \\
\quad
$^2$Fudan University
\quad
$^3$Northwestern Polytechnical University
\smallskip
\\ 
\small{\texttt{\{shujianjiao,wanglei94,xiaojinzhang,lemuria\_chen\}@hust.edu.cn}}\\
\small{\texttt{windskileoli@gmail.com}~~} 
\small{\texttt{jiajiepeng@nwpu.edu.cn}~~} 
\small{\texttt{zywei@fudan.edu.cn}~~} 
}

\begin{document}
\maketitle
\def\thefootnote{*}\footnotetext{Equal contribution. \textsuperscript{\Letter}Corresponding authors.}
\def\thefootnote{\arabic{footnote}}
\begin{abstract}

Proteins are essential to life's processes, underpinning evolution and diversity. Advances in sequencing technology have revealed millions of proteins, underscoring the need for sophisticated pre-trained protein models for biological analysis and AI development. Facebook's ESM2, the most advanced protein language model to date, leverages a masked prediction task for unsupervised learning, crafting amino acid representations with notable biochemical accuracy. Yet, it lacks in delivering functional protein insights, signaling an opportunity for enhancing representation quality.Our study addresses this gap by incorporating protein family classification into ESM2's training.This approach, augmented with  Community Propagation-Based Clustering Algorithm, improves global protein representations, while a contextual prediction task fine-tunes local amino acid accuracy. Significantly, our model achieved state-of-the-art results in several downstream experiments, demonstrating the power of combining global and local methodologies to substantially boost protein representation quality.
\end{abstract}

\section{Introduction}
Proteins are involved in almost all the life activities of organisms, and the study of their sequences, structures, characteristics, and roles is a major area of research in the life sciences in the post-genomic era~\cite{papin2003metabolic}.A protein sequence can be thought of as a string of amino acid letters. The residues, structural domains, and families of amino acids that make up a protein resemble words, phrases, and sentences in human language. Therefore, machine learning methods developed for natural language and other sequences are well suited to the task of predicting proteins~\cite{ofer2021language}.Most sequence-based language models [e.g., BERT ~\cite{devlin2018bert}, XLNet ~\cite{yang2019xlnet}, ELECTRA ~\cite{clark2020electra}] are designed to process natural language (with a bias towards English).Considering the notable parallels between protein sequences and the structure of natural language, employing natural language processing~\cite{chowdhary2020natural} (NLP) techniques to analyze protein sequences emerges as a logical approach.

Currently, ESM-2~\cite{lin2022language}, developed by Facebook, is recognized as the most extensive protein sequence language model to date, featuring a sophisticated architecture with 48 layers and over 15 billion parameters. This groundbreaking model is trained on an expansive dataset comprising up to 250 million protein sequences sourced from Uniparc~\cite{leinonen2004uniprot}, which encompasses 86 billion amino acids~\cite{ng2006predicting}. The dataset's vast scale mirrors the extensive text corpora used in developing large-scale neural networks for natural language processing, highlighting the model's unmatched breadth and depth. Leveraging this comprehensive base, ESMFold emerges as an innovative 3D protein structure prediction tool, leveraging ESM-2's insights with just a single sequence input to significantly speed up predictions. In cases where it deeply understands sequences, ESMFold achieves atomic-level accuracy, matching or even surpassing leading models like AlphaFold2~\cite{jumper2021highly} and RoseTTAFold~\cite{baek2021accurate}. This efficiency in generating swift, precise predictions from single inputs showcases that a more profound grasp of sequences leads to a better understanding of protein structures.

Challenges accompany ESM-2 and its derivatives like ESMFold, primarily in their reliance on statistical analysis of amino acid compositions to generate amino acid representations. This approach, focused on predicting missing amino acids within contexts, proves adequate for classifying protein families and identifying remote homologies but falls short in capturing the full functional complexity of proteins. Consequently, the unsupervised training method yields protein representations that lack interpretability, underscoring the insufficiency of statistical properties alone to fully elucidate protein functions. Moreover, since the computational complexity is the square of the sequence length and the length of protein sequences is much larger than the text length, and since ESM2 and ESMFold require a large amount of computational resources, the large number of parameters, and the complexity of the computation pose obstacles for researchers who do not have high-performance computing facilities. Additionally, while ESMFold offers promising capabilities in predicting protein structures from single sequences, it struggles with proteins exhibiting complex folding patterns or those necessitating interactions with other molecules for accurate structural and functional modeling, indicating the potential need for integrating additional sequence or molecular information for more precise predictions.\par
In general, our contributions are of three-folds:

\begin{itemize}
\setlength\itemsep{-0.5em}
\item We have fused graph pre-training with masked language modeling to refine the ESM2 model, achieving unparalleled performance in protein-centric tasks beyond ESM2.
 \item Our proposed Community Propagation-Based Clustering Algorithm is a novel, resource-efficient training method for graph neural networks.
 \item We provide a detailed demonstration of the role of the graph network asynchronous information propagation algorithm in the pre-training tasks of protein sequences.
\end{itemize}

\begin{figure*}[htbp]
\centering
\includegraphics[scale=0.5]{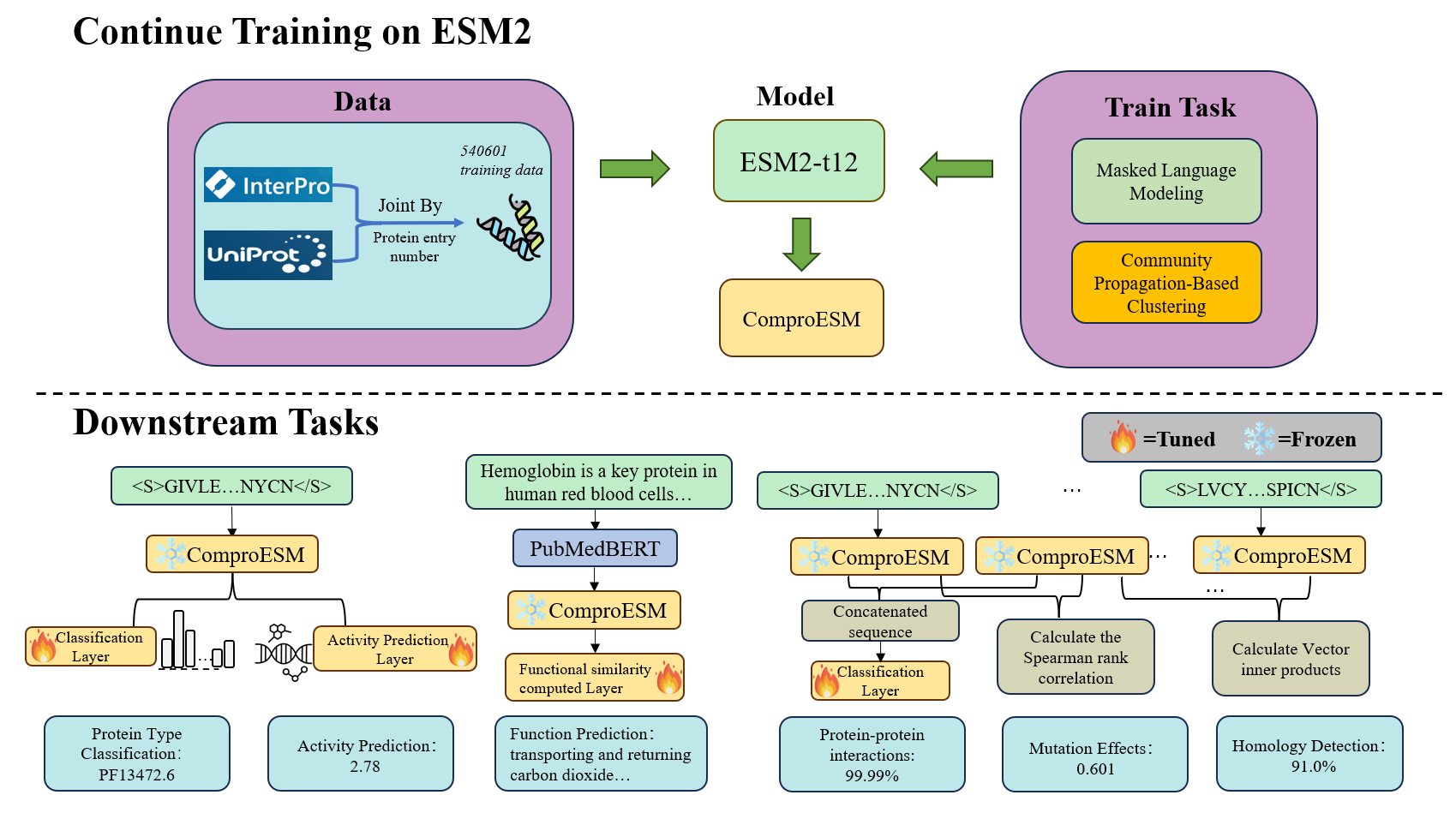}
\caption{The overview of our work. }
\label{Model}
\end{figure*}

\section{Related Work}

\subsection{Protein Large Language Models}

The technology of language models is currently very popular and being explored for application across a variety of professional domains, including healthcare~\cite{bao2023disc}, finance~\cite{chen2023benchmark}, industry~\cite{gu2024anomalygpt}, education~\cite{dan2023educhat}, and the judiciary~\cite{yue2023disc}. In the healthcare sector, there is a particular focus on AI-based intelligent diagnostics~\cite{zhong2022hierarchical,chen2023benchmark,chen2023dxformer,fan2024ai}, with primary modalities including text, medical imaging, and medical examinations~\cite{gu2020automatic}. As research delves deeper, it has become evident that understanding the microscopic information, such as genetics and proteins, is crucial for gaining a more profound comprehension of disease processes. 

Proteins, with their sequences of amino acids, exhibit a remarkable resemblance to natural language, where residues and structural domains can be likened to letters and words. This similarity has sparked interest in developing AI methodologies tailored for protein analysis. By leveraging the parallels between the intricate 'language' of proteins and human language, researchers are now turning their attention to AI-driven approaches to better understand and predict protein sequences, which are fundamental to the mechanisms of life. The development of specialized Protein Large Language Models (Prot-LLMs) has significantly advanced with models categorized into encoder-only, decoder-only, and encoder-decoder architectures, each tailored for different protein research applications. Encoder-only models are optimized for predicting protein functions or properties, while decoder-only models focus on protein sequence generation. Central to encoder-only Prot-LLMs is the Transformer encoder architecture, which efficiently translates protein sequences into concise vector representations, crucial for identifying unique protein patterns. Prominent pre-trained protein sequence encoders include ProteinBert~\cite{brandes2022proteinbert}, ProtTrans~\cite{elnaggar2021prottrans}, PMLM~\cite{he2021pre}, ProtFlash~\cite{wang2023deciphering}, ProtNPT~\cite{notin2023proteinnpt}, ESM1b~\cite{rives2021biological}, ESM-1v~\cite{meier2021language}, and ESM-2~\cite{lin2022language}. The integration of techniques such as Multiple Sequence Alignment (MSA) with ESM-MSA-1b (MSA Transformer)~\cite{rao2021msa} has been instrumental in breakthroughs like AlphaFold2 and AlphaMissence~\cite{cheng2023accurate}.Incorporating 3D structural information has marked a significant progression, exemplified by models like ESM-GearNet~\cite{zhang2023enhancing}, SaProt~\cite{su2023saprot}, LM-GVP~\cite{wang2022lm}, and PromptProtein~\cite{wang2022multi}, indicating a shift towards structure-aware pretraining to refine protein representation. Efforts to enhance encoder architectures and training methodologies include self-supervised learning through masked language modeling (MLM) tasks, effectively reconstructing corrupted tokens from contextual cues, with the ESM series leveraging the Transformer encoder architecture, akin to BERT and RoBERTa~\cite{liu2019roberta}, to predict protein structure and function.

\subsection{Advancements in ESM Models}
The field of protein prediction has witnessed transformative advancements through the evolution of Evolutionary Scale Modeling (ESM). A significant leap forward was achieved with the adaptation of the ESM1b model, which now provides highly accurate variant effect predictions for an extensive array of over 40,000 protein isoforms, markedly surpassing the capabilities of conventional methodologies. The introduction of the ESM-2 model represents another milestone, featuring an exponential increase in model parameters from 8 million to an impressive 15 billion. This upscaling has been instrumental in achieving unprecedented accuracy in protein structure prediction, demonstrating the critical role of model complexity in computational protein analysis. Furthermore, the application of ESM models in specialized tasks, such as MHC peptide binding prediction~\cite{hashemi2023improved} and single-sequence protein structure prediction~\cite{wang2022single}, showcases the models' adaptability and effectiveness across diverse research applications. The release of cutting-edge pre-trained models and codes~\cite{esm_facebookresearch} specifically designed for protein design, coupled with the recent updates to the ESM Metagenomic Atlas, underscores the continuous efforts toward refining the predictive accuracy and utility of ESM models. Moreover, the integration of LR-ESM in the NetGO 3.0~\cite{wang2023netgo} framework for advanced large-scale functional prediction exemplifies the broadening scope of ESM models in providing critical insights into the vast expanse of uncharacterized proteins. Collectively, these advancements not only enhance our understanding of protein functions and interactions but also pave the way for novel discoveries in protein science.
\section{Method}
The masking language task is crucial for the model to discern each amino acid's biochemical traits, while Community Propagation-Based Clustering connects these sequences to the protein's structure and function based on the statistical nature of amino acids. The masked language model deciphers semantics at the residue level, and the clustering approach interprets the entire protein sequence. Together, they meld local and global insights, greatly enhancing protein representation and capturing the intricate relationship between a protein's structure, function, and statistics.Community Propagation-Based Clustering Algorithms, inspired by hierarchical clustering, enable nuanced classification of proteins into detailed families or broader superfamilies, establishing hierarchical relationships. This mimics the organization of communities and towns, where synchronizing tasks among members—akin to information exchange within a network—is managed through regular updates to a central system. This model of synchronization is analogous to the information propagation on a graph, streamlining the integration and coordination of work progress.\par 
\subsection{Community Propagation-Based Clustering }
\begin{figure}[htbp]
\centering
\includegraphics[scale=0.27]{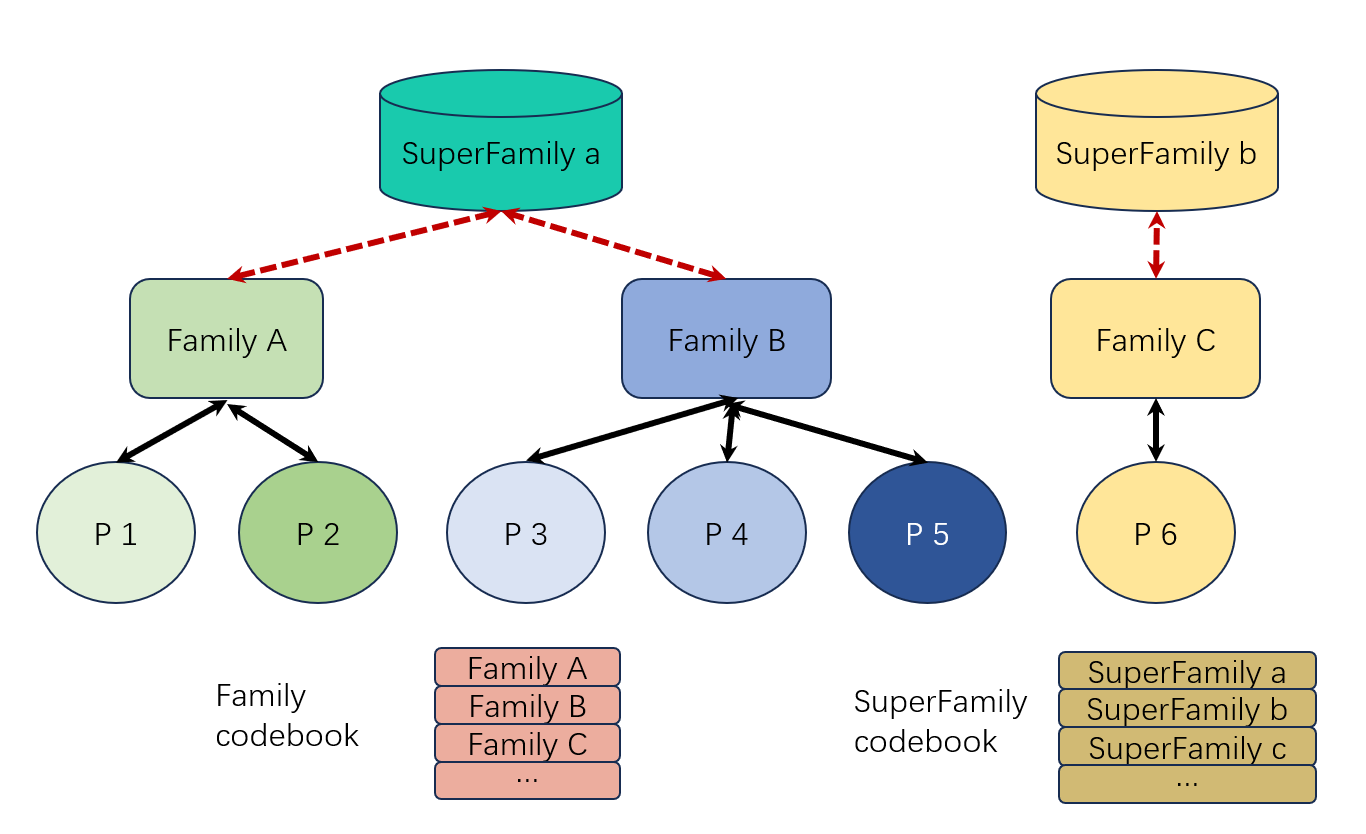}
\caption{Direction of information flow for Community Propagation-Based Clustering Algorithm}
\label{community}
\end{figure}
In Figure \ref{community}, P1 and P2 are two proteins in family A, and P3 and P4 are two proteins in family B. Together they belong to a superfamily. The black realisation arrows represent the information exchange between the protein nodes and the family nodes they belong to, and this exchange should be executed immediately along with the update of the protein nodes. The red dotted line represents the information exchange between the family node and the superfamily node, which should be time-delayed, i.e., when the update on the family accumulates a certain number of steps before it is performed.\par
What needs to be considered in the Community Propagation-Based Clustering Algorithm is how to design the initial representation of each node and how information is propagated between nodes. For the initial representation of a protein node, this paper uses the average pooling of the protein representation model for the representation of each amino acid residue output as the initial representation of the protein node:
\begin{equation}
P = \frac{1}{l}\sum_{i=1}^l{H_i}, 
\end{equation}
where $l$ represents the input sequence length and $H_i$ represents the representation vector of the $i$th amino acid in the last layer of the protein representation model. In this paper, we use randomly initialised vectors $F\in\mathbb{R}^d$ and $SF\in\mathbb{R}^d$ as the initial representations of family and superfamily nodes, respectively.

In this paper, link prediction is used as the information propagation algorithm. Specifically, for the input protein representation $\{P_1, P_2, ... , P_N\}$, $N$ being the batch size, and the corresponding family node representation $\{F_1, F_2, ... , F_N\}$ and the superfamily node representation $\{SF_1, SF_2, ... , SF_N\}$. The score function $S_F(P_i, F_j)$ is used to evaluate the likelihood of establishing a link between protein node $i$ and protein family node $j$. The score function $S_{SF}(P_i, SF_j)$ is used to evaluate the possibility of establishing links between protein nodes and superfamily nodes, and in practice this paper adopts the vector inner product as the score function. It should be ensured that the distribution of the output of the score function conforms to the distribution of the actual graph.

In this paper, ${\rm Softmax}$ is used to measure the distribution of the score function. Specific:
\begin{equation}
\mathcal{L}_{family} = \frac{1}{N}\sum_{i=1}^N{{\rm -log Sm}(S_F(P_i, F_i))},
\end{equation}
\begin{equation}
\mathcal{L}_{superfamily} = \frac{1}{N}\sum_{i=1}^N{{\rm -log\, Sm}(S_{SF}(P_i,SF_i))},
\end{equation}
where ${\rm Softmax}()$ measures the distribution of samples in a batch. The loss representation of the final Community Propagation-Based Clustering Algorithm is:
\begin{equation}
\mathcal{L}_{\rm commu}=w\mathcal{L}_{family}+(1-w)\mathcal{L}_{superfamily},
\end{equation}

where $w$ represents the weight magnitude of the two losses.
The Community Propagation-Based Clustering Algorithm proposed in this paper addresses how data with multiple category labels can be incorporated into a pre-trained language model by means of graph pre-training. Although the external manifestation of the algorithm is the propagation of information on a graph, it essentially addresses the problem of hierarchical clustering of data from different categories.

Clustering algorithms, such as hierarchical clustering~\cite{murtagh2012algorithms} and k-means clustering~\cite{hartigan1979algorithm}, segment data into classes based on feature similarity, aiming to maximize intra-class similarity and minimize inter-class similarity. In the realm of protein families, this translates to grouping proteins into superfamilies and families, where each family represents a fine-grained cluster and each superfamily a coarse-grained one. This research treats each family as a precise clustering focus, reflecting the need for protein representations to encapsulate the familial information. Hence, we introduce the Community Propagation-Based Clustering Algorithm for a supervised approach to clustering protein sequences, aligning the sequence representation with its familial heritage.A detailed exposition on the algorithm's interpretability is provided in the appendix\ref{7}.

\subsection{Masked Language Modeling}
Intertwined with clustering, the masked language task equips the model to discern individual amino acid biochemical properties. The Community Propagation-Based Clustering Algorithm then takes this a step further, forging connections between amino acid sequences and the protein's structural and functional attributes. These attributes are deeply linked to the statistical properties of amino acids. Together, these twin objectives—masking language tasks and community propagation—create a synergy, reinforcing each other to refine the protein representation quality.
In our methodology, we harness the BERT model's architecture, renowned for its multiple stacked Transformer Blocks, as the encoder for protein sequences. To achieve our dual goals, we deploy two distinct fully connected neural networks to project the encoded protein vectors into separate representation spaces for the masked prediction task and the Community Propagation-Based Clustering Algorithm. Each Transformer Block in the BERT framework features a multi-head self-attention mechanism, recalculating token representations by considering contextual token relationships. Multi-heads permit multiple attention computations, capturing diverse sequence characteristics, followed by concatenation and mapping back to the original input size through a perceptron. The forward propagation network—a classic two-layer fully connected network—serves as an intermediary, transforming representations between layers.

Each token vector encoded by the BERT model will have its loss computed by the mask prediction task and the Community Propagation-Based Clustering Algorithm, respectively, in order to run the stochastic gradient descent algorithm to update all network parameters of the model. Specifically, for the mask prediction task the loss function is computed as:
\begin{equation}
\mathcal{L}_{\rm MLM} = \mathbb{E}_M\sum_{i\in M}-{\rm log}p(x_i|x_{/M}),
\end{equation}
For each sequence $x=\{x_1, x_2, ... , x_n\}$, a set of tokens $M$ to be masked is sampled, and the actual amino acid at each position $i$ is replaced with the mask symbol <mask>. For each masked position, the loss function is set to the negative logarithmic value of the probability of predicting the correct amino acid $x_i$, taking the sequence $x_{/M}$ as the context of the protein sequence with the masked portion removed. Intuitively, in order to predict the masked position, the model must identify the correlation between the masked site and the unmasked portion of the sequence. The probability is calculated as:
\begin{equation}
p(x_i|x_{/M})={\rm Softmax}({W_M}_j \cdot z_{li}),
\end{equation}
where $W_M\in\mathbb{R}^{d \times V}$ is the vocabulary prediction matrix used to map the size of the representation vector of each token to the size of the vocabulary list, $V$ is the size of the vocabulary list, and ${W_M}_j$ stands for the vector of prediction matrices corresponding to the $j$ word (the correct word corresponding to $x_i$) in the vocabulary list.

For the Community Propagation-Based Clustering Algorithm this paper uses the link prediction task as the goal, using the average of the token representations at each location as the final representation vector of the protein sequence i.e:
\begin{equation}
P = \frac{1}{l}\sum_{i=1}^l{z_{li}},
\end{equation}

where $l$ represents the input sequence length and $z_{li}$ represents the representation vector of the $i$th amino acid output from the last layer of BERT. In this paper, we use the randomly initialised vectors $F\in\mathbb{R}^d$ and $SF\in\mathbb{R}^d$ as the initial representations of family and superfamily nodes, respectively. Assume that all protein sequences involved in a single batch calculation are denoted as $\{P_1, P_2, ... , P_N\}$, $N$ being the batch size, and the corresponding family node representation for each protein $\{F_1, F_2, ... , F_N\}$ and the superfamily node representation $\{SF_1, SF_2, ... , SF_N\}$. The score function $S_F(P_i, F_j)$ is used to evaluate the likelihood of establishing a link between protein node $i$ and protein family node $j$. The score function $S_{SF}(P_i, F_j)$ is used to evaluate the possibility of establishing links between protein family nodes and superfamily nodes, and in practice this paper adopts the vector inner product as the score function. It should be ensured that the distribution of the score function output matches the distribution of the actual graph.

The final model loss is the sum of the two optimisation objective losses:
\begin{equation}
\mathcal{L}_{\rm model} = \mathcal{L}_{\rm MLM} + \mathcal{L}_{\rm commu}.
\end{equation}
The specifics concerning the masked language model and the Community Propagation-Based Clustering are elaborated in the appendix\ref{8}.
\section{Continue Training on ESM2}\label{Ex}
\begin{table*}[htbp]
\centering
\caption{Experimental validation of the model on downstream tasks.}
\label{metrics}
\resizebox{\textwidth}{!}{
\begin{tabular}{lcccccc} \toprule
Model & Family Class. & Mut. Effects & Act. Pred. & Interact. & Func. Pred. & Homol. Det. \\ \midrule
ESM2-12layer & 27.51\% & 0.514 & 0.040 & 99.68\% & 90.38\% & 87.8\% \\
ESM2-6layer & 20.55\% & 0.423 & 0.043 & 99.58\% & 88.39\% & 81.3\% \\
TinyBert & 13.30\% & 0.178 & 0.075 & 99.17\% & 88.09\% & 76.5\% \\
\textbf{OurModel} & \textbf{31.08\%} & \textbf{0.601} & \textbf{0.038} & \textbf{99.99\%} & \textbf{91.47\%} & \textbf{91.0\%} \\
 \midrule
Ablation Experiment & & & & & &\\
 \midrule
only MLM & 27.93\% & 0.601 & 0.042 & 99.75\% & 90.27\% & 88.1\% \\
only Graph & 30.05\% & 0.132 & 0.052 & 99.69\% & 91.0\% & 89.2\% \\ \bottomrule
\end{tabular}
}
\end{table*}
\subsection{Pre-training Dataset}
We obtained the classification data and corresponding amino acid sequences of proteins from two databases, InterPro (\url{https://www.ebi.ac.uk/interpro/})~\cite{paysan2023interpro} and UniprotKB (\url{https://www.uniprot.org/})~\cite{ boutet2016uniprotkb} are two databases to obtain the classification data and corresponding amino acid sequences of the proteins, respectively. interPro classifies each protein into a family and provides predictions of protein structural domains and sites of importance in the protein sequence to analyse the protein function. To classify proteins in this way, InterPro uses prediction models provided by several different databases that make up the InterPro consortium (called member databases). By combining protein features from these member databases into one searchable resource, their individual strengths are exploited to generate a powerful integrated database and diagnostic tool. 

UniProtKB contains the functional information of the proteins and preserves the amino acid sequences of the proteins.UniProtKB consists of two parts:
\begin{itemize}
    \item The information in this part of the UniProtKB/Swiss-Prot-Database is manually annotated and reviewed, and is therefore of high quality and non-redundant.
    \item UniProtKB/TrEMBL-The information in this part of the database is computationally annotated by algorithms and is not reviewed, thus providing a high level of coverage of the proteome.
\end{itemize}
The protein classification data from InterPro and the protein sequence data from UniproKB are finally joined by protein entry numbers to obtain all the training data.

Considering the limitations of local computing resources and the quality of the data, all the proteins from UniProtKB/Swiss-Prot database are finally extracted in this paper, which are all These proteins are all manually annotated and thus representative and highly reliable.

A total of 568744 proteins were obtained from UniProtKB
/SwissProt, and all the obtained proteins were queried for the corresponding protein interfaces in InterPro, and finally 540601 training data were obtained. For each protein, the complete amino acid sequences of all families and superfamilies were recorded. It should be noted that some proteins may lack the corresponding superfamily or family classification information, but they must have one of the two categories, in this case, we will randomly sample to fill in the missing items, for example, if a protein lacks a family category, the model will randomly select one of the superfamily categories as the family category each time it reads the sample. Detailed statistics of the data are given in the table\ref{stastic}.

\begin{table}[htbp]
\caption{Statistics of the dataset.}
\centering
\small
\begin{tabular}{lr} \toprule
Statistics & Values \\ \midrule
Protein samples count & 540,601 \\
Family categories count & 17,132 \\
Superfamily categories count & 3,189 \\
Family memberships per protein & 1.23 \\
Family memberships per protein & 1.43 \\
Average length of amino acid sequences & 367.01 \\ \bottomrule
\end{tabular}
\label{stastic}
\end{table}

\subsection{Training Details}
A 12-layer ESM2 model was used as the protein representation model in the pre-training experiments, which was subjected to knowledge distillation using Tinybert to generate the student model. For the mask prediction task, 15\% of the amino acid residues in the protein sequence are randomly selected for replacement, of which 80\% are replaced with <mask>, 10\% are replaced with a random representation, and the remaining 10\% are kept unchanged. For Community Propagation-Based Clustering Algorithm, the weights of family score loss and superfamily score loss were 0.2 and 0.8, respectively. an AdamW was used as the optimiser with a learning rate of 5e-5 .

\subsection{Validation of Protein Representation}
\begin{figure}[htbp]
\centering
\includegraphics[scale=0.16]{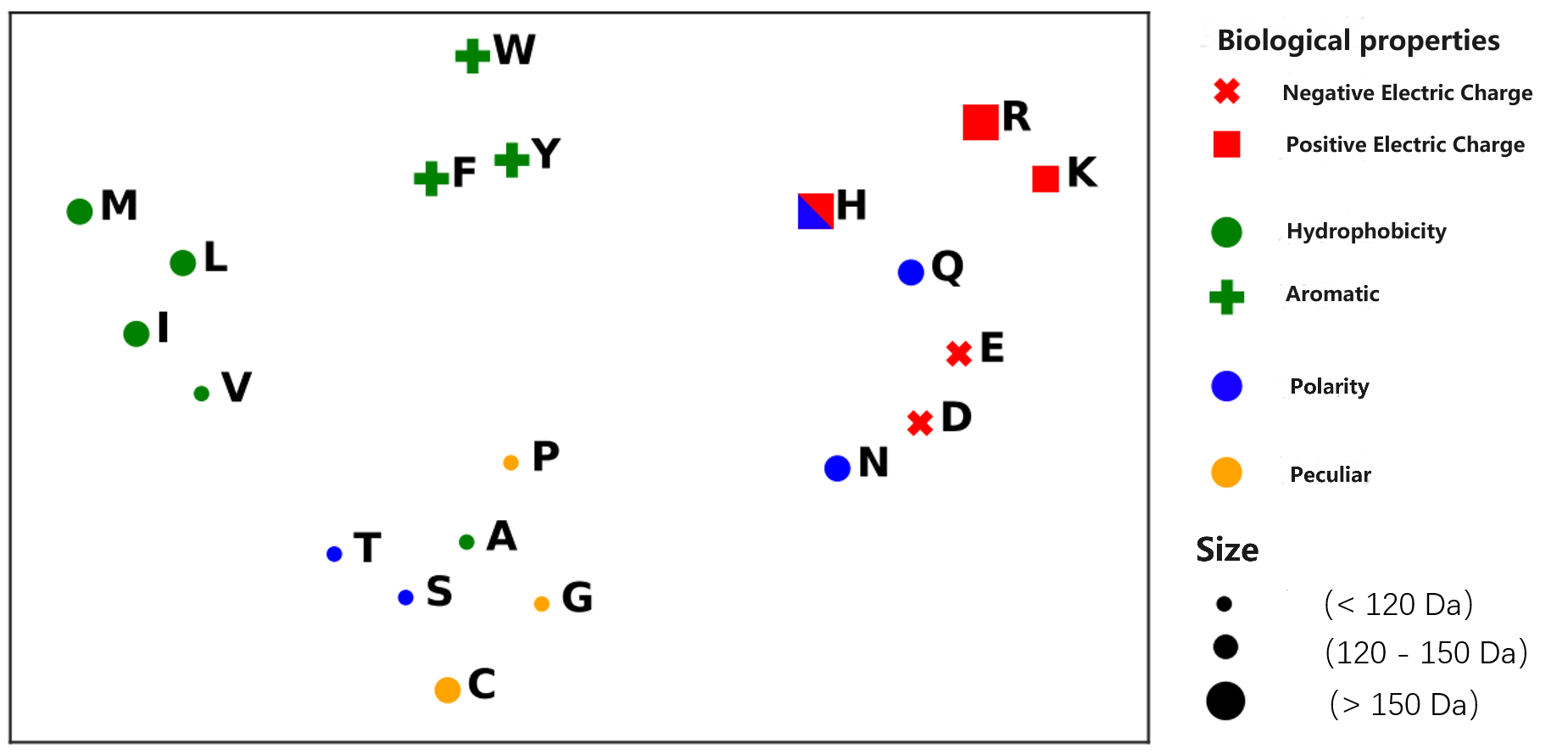}
\caption{Model prediction layer representation of the biochemical properties of the embedded amino acids.}
\label{AA1}
\end{figure}

The visualisation method demonstrates whether the stability and functional properties possessed by a protein can be deduced from the representation of the model. The stability of a protein is mainly characterised by the biochemical properties of the amino acid residues that make up the protein, whereas the function of a protein is mainly reflected in the proximity of representations of proteins with similar structure and function.In order to investigate whether the model has learnt to encode physicochemical properties in the representation of amino acid residues, this paper uses TSNE to downscale the parameters of the final prediction layer of the network so that they can be plotted onto 2D coordinates. In Fig. \ref{AA1}, it is found that the space can clearly distinguish the boundaries between amino acids with different biochemical properties. These biochemical properties include hydrophobicity, polarity, and aromaticity, as well as molecular size and electrode properties, all of which affect the interchangeability of amino acid residues at a site in a protein, i.e., the stability of the protein structure.

\begin{figure}[htbp]
\centering
\includegraphics[scale=0.33]{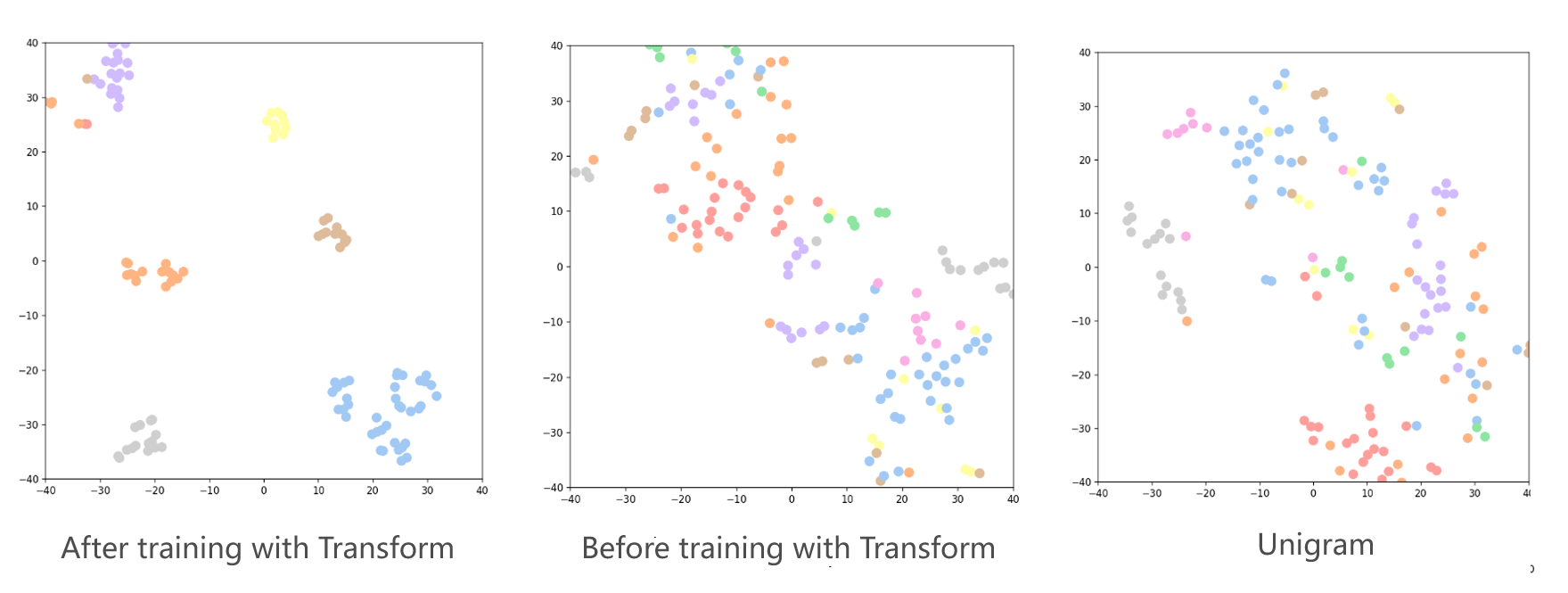}
\caption{Protein representation after TSNE reduction}
\label{tsne_rep}
\end{figure}

A representation of each amino acid residue can be obtained through the final layer of the model, and the final protein can be obtained by averaging these representations. These protein representations project a sequence to a point in space, thus different protein sequences can be distinguished. In this paper, the distribution of homologous genes in space can be studied intuitively in two-dimensional coordinates by TSNE dimensionality reduction. In addition, Principal Component Analysis (PCA) was applied to recover the main direction of change in the representation, and four homologous genes in four species were selected to find the direction of change. In order to exclude the statistical properties of the number of amino acids itself in the amino acid sequence, this paper also introduces the protein representation before untrained and the protein representation using the unitary bag-of-words model. Figure \ref{tsne_rep} shows that the modelled protein representations encode structural and functional information about the proteins thereby allowing for a significant aggregation effect of protein representations within the same protein family. 

\section{Downstream task experiments}
\subsection{Downstream task introduction}
To evaluate the effectiveness of the protein representation model post-pretraining, we established the following downstream tasks based on six different protein aspects: Protein Classification, Protein Activity Prediction, Protein-Protein Interaction, Protein Function Prediction, Mutation Effects, and Homology Testing. For task validation, datasets specific to each task were procured, which were not involved in the training of the protein representation model. The model developed in this study solely generates a numerical vector replete with structural-functional information from the input protein sequences, which is then utilized for the respective downstream tasks. For tasks requiring multiple protein sequences (Protein-Protein Interaction, Mutation Effects, Homology Testing), we adopted an approach of training multiple protein sequences over several iterations. Tasks such as Protein Classification, Protein-Protein Interaction, and Protein Function Prediction necessitate the construction of a multi-layer fully connected neural network and subsequent fine-tuning on a training set. It is important to note that the parameters of the protein representation model proposed in this paper remain constant during the experiments of these downstream tasks. Detailed experimental procedures are provided in the appendix(from appendix\ref{1} to appendix\ref{6}).
\subsection{Enhanced Performance in Protein Representation Tasks}
The table highlights optimal performances in bold, underscoring that ComproESM—leveraging both mask prediction tasks and the graph network asynchronous information propagation algorithm—surpasses the ESM2 model in a range of downstream tasks. Given ComproESM's integration of protein family data, which encapsulates structural and functional protein information, enhancements in family classification, function prediction, and homology detection are anticipated. Moreover, ComproESM outperforms the ESM2 model in tasks requiring a comprehensive understanding of the biochemical properties of amino acid residues, such as mutation effects, activity prediction, and protein-protein interactions. Ablation studies focusing individually on these tasks further suggest that while targeting specific tasks can yield improvements, a combined approach garners optimal results across all downstream tasks.
Considering both ESM2 and our proposed model are trained on mask prediction tasks, we partitioned data into a training set and test set at a 9:1 ratio. Performance on the mask prediction task is evaluated using the Exponential Cross Entropy Score (ECE), where a perfect model scores 1, and a uniform random prediction across 25 amino acids scores 25. Lower ECE scores indicate superior mask prediction performance. As shown in Table \ref{MLM_result}, our model slightly underperforms the sole MLM model trained on the same dataset, attributed to the incorporation of graph link prediction tasks altering the original protein representation space by integrating protein family data.Further validation of the changes induced by the protein representation is detailed in the appendix\ref{9}.

\begin{table}[htbp]
\caption{Mask prediction task.}
\centering
\small
\begin{tabular}{lcc}
\hline
Model & Parameter Count & ECE \\
\hline
Ideal value & & 1 \\
Uniform Distribution & & 25 \\
\hline
4-gram & & 19.21 \\
LSTM-small & 28.4M & 15.13 \\
LSTM-Large & 113.4M & 13.97 \\
ESM2-12layer & 35.8M & 5.80 \\
OurModel & 35.8M & 5.78 \\
only MLM & 35.8M & 5.62\\
\hline
\end{tabular}
\label{MLM_result}
\end{table}

\subsection{Protein Classification Model Analysis}
In order to analyse the advantages of the model in this paper on the protein classification task even further, the 10 most common protein families were selected from the dataset PF13649.6, PF00560.33,PF13508.7,PF06580.13,PF02397.16,
PF00677.17,PF01035.20,PF02417.15,PF13472.6,
PF00684.19.60 samples of proteins were sampled for each category, of which 50 samples were used for training and 10 samples for testing. The results after 20 epochs of iteration under the same experimental setup are shown in Table \ref{task_class_result}, and it can be found that the accuracy of the model proposed in this paper is much larger than that of the ESM2 model in the training set, which indicates that the protein representation given by the model proposed in this paper has a better decision boundary between the different protein family classes, and so the model can be in training to find this boundary quickly. The confusion matrix of the prediction results is shown in Figure \ref{confusion_pic}, which shows that the ESM2 model has a lower prediction accuracy for the proteins in the protein family categories PF13649.6, PF13508.7, and PF13472.6, whereas there is no such bias in the model proposed in this paper, which indicates that the protein representations given in the proposed model are more consistent with the protein family categories than those given in the ESM2 model. protein representation has a more direct correlation with the protein family categories.
\begin{table}[htbp]
\caption{Comparison of results on the protein classification task}
\centering
\small
\begin{tabular}{lcc}
\hline
Models & Training Sets & Test Sets \\
\hline
ESM2-12layer & 38.46\% & 80.00\% \\
OurModel & 78.24\% & 94.00\% \\
\hline
\end{tabular}
\label{task_class_result}
\end{table}

\section{Conclusion}

In this paper, we propose ComproESM, and demonstrate the efficacy of integrating protein family classification information with the ESM2 model's training process, thereby enhancing the model's capability to accurately represent global protein structures and predict local amino acid residues. Our innovative Community Propagation-Based Clustering Algorithm, when combined with the masking algorithm, has significantly advanced the model's performance across a spectrum of downstream tasks, far surpassing the capabilities of the ESM2 model with a comparable number of parameters.These findings not only underscore the potential of tailored algorithmic strategies in understanding complex biological data but also pave the way for future research in protein sequence analysis and functional prediction. Through this work, we contribute to the broader field of computational biology by providing a robust framework for protein representation that is sensitive to both the structural and functional nuances of proteins. 


\section*{Limitation}

Although this paper achieves optimal performance on various protein downstream tasks, due to the limitation of experimental resources, we does not use the mask prediction task and the Community Propagation-Based Clustering Algorithm to train directly on the randomly initialised Transformer model, but uses the 12-layer model of ESM2 for initialisation. The representation space of the final model trained by this method is highly offset, which means that the model only converges to a locally optimal solution. Meanwhile, the training data taken in this paper only has 500,000 proteins, which is still too small compared to the hundreds of millions of proteins stored in protein databases. In the future, this paper will train the model in this paper from scratch on larger protein sequence data, so as to obtain a higher quality protein representation model.

\section*{Ethics Consideration}

Given the profound potential impact of this research on the life sciences and medical fields, several ethical considerations must be taken into account. Firstly, ensuring data privacy is critical, particularly if any human-derived protein sequences are included in the dataset; researchers must handle such data responsibly and with consideration for consent and confidentiality. Additionally, the accessibility of the developed tools and methods should be addressed to prevent exacerbating inequalities in research capabilities across different institutions or countries, especially considering the computational resources required for high-performance models like ESM-2. Moreover, it is imperative to transparently communicate both the capabilities and limitations of the model to avoid misinterpretation or overreliance on its predictions, which could have significant consequences for downstream applications such as drug discovery or personalized medicine. Finally, there should be a commitment to open science by making the methodologies, source code, and potentially the trained model accessible to other researchers, fostering collaboration and accelerating advancements while respecting intellectual property rights.


\bibliography{custom}

\appendix

\label{sec:appendix}

\section{Interpretability of Community Propagation-Based Clustering Algorithms}\label{7}
\begin{figure}[htbp]
\centering
\includegraphics[scale=0.36]{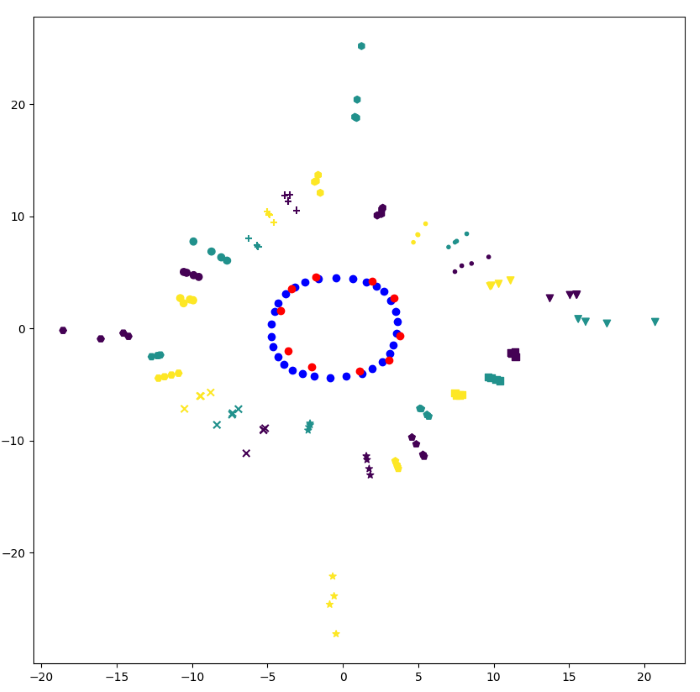}
\caption{Community Propagation-Based Clustering Algorithm on random initialisation vectors}
\label{propagation}
\end{figure}

In the Community Propagation-Based Clustering Algorithm, the data input to the model is a set of samples $\{P_1, P_2, ... , P_N\}$, where $N$ is the batch size, and each sample corresponds to the family category representation $\{F_1, F_2, ... , F_N\}$ and the superfamily category representation $\{SF_1, SF_2, ... , SF_N\}$. For the $i$th sample $P_i$ of these, $\rm Softmax(Z_i)$ is required to be equal to 1. That is, the score $Z_i$ is required to be as large as possible and the other scores as small as possible. Since $Z = w Z_F + (1-w) Z_{SF}$, i.e., the scores $Z_F$ and $Z_{SF}$ are required to be as large as possible. Take the calculation of $Z_F$ as an example:
\begin{equation}
Z_F = P \cdot F.
\end{equation}
where $P = \{p_1, p_2, ... , p_d\}, F=\{f_1, f_2, ... , f_d\}$, then $Z_F$ can be expressed as:
\begin{equation}
Z_F = \sum_{j=1}^{d}{p_j \times f_j},
\end{equation}
And if you use the Euclidean distance measure between $P, F$, the distance distance is given by:
\begin{equation}
S = \frac{\sum_{j=1}^{d}{(p_j-f_j)^2}}{|P||F|}, \frac{\sum_{j=1}^{d}{(p_j-f_j)^2}}{|P||F|},
\end{equation}
\begin{equation}
\frac{|P|^2+|F|^2-2Z_F}{|P||F|}, 
\end{equation}
Assuming that the stretching of the vectors does not affect the properties of the vectors themselves, it can be assumed that $|P|=|F|=1$, and thus the distance $S$ can be expressed as:
\begin{equation}
S = 2 - 2Z_F.
\end{equation}
It follows that ${\rm max}(Z_F)$ is equivalent to ${\rm min}(S)$, in other words the optimisation objective treats the category representation as a clustering centre, and the purpose of the loss function is to bring in closer the distance between nodes within the class and the centre of the class, and to bring out further the distance between nodes outside the class and the centre of the class. The parameters $w$ and size are used to control the coverage of different granularity categories in the representation space.
In order to demonstrate the clustering effect of Community Propagation-Based Clustering Algorithm in a more graphical way, in this paper, 120 random 2-dimensional vectors are initialised using a normal distribution, and each vector corresponds to a fine-grained category and a coarse-grained category, with a total of 30 small categories and 10 large categories. The backpropagation algorithm was used on this randomly generated dataset using Community Propagation-Based Clustering Algorithm until convergence, and finally the representations of all the 2-dimensional vectors were presented on the axes (as shown in Fig. \ref{propagation}).

In the graph \ref{propagation}, all red nodes represent coarse-grained cluster centres and all blue nodes represent fine-grained cluster centres. The density of the red nodes is less than that of the blue nodes due to the weight parameters $w_f$ and $w_s$. The rest of the nodes are all nodes in the dataset, nodes belonging to the same coarse-grained level are represented using the same shape, and nodes of the same colour under the same shape represent their same fine-grained category. From the figure, it can be seen that Community Propagation-Based Clustering Algorithm achieves a significant clustering effect and the sample distribution is in accordance with the assumption above, i.e., the stretching of the vectors does not affect the nature of the vectors themselves.\par

\section{Mask Language Model Setting Details}\label{8}
The input used by the model is a protein sequence, which is a string of letters where each letter represents a particular amino acid. Also for each protein sequence there is a family class and superfamily class to which the protein belongs, and these labels will be used for subsequent loss function calculations. The model will first obtain a token sequence from the input protein sequences (each letter, i.e., an amino acid, is a token) and add special identifiers <bos> and <eos> at the beginning and the end of the sequence, respectively, and use a masking technique to blur the input so as to force the model to derive the missing token based on the context. 15\% of the tokens will be randomly selected and used with the identifier <mask>. A random 15\% token is selected and replaced with the identifier <mask>. Note that in order to improve the model's performance, the strategy of replacing the original token with <mask>, replacing it with a random token, and keeping the original token unchanged is implemented in a ratio of 8:1:1 for all the replaced words. The inputs to the model are $\{x_1, x_2, ... , x_n\}$ where $x_i\in\mathbb{R}$, the first layer of the BERT model is the word encoder, which is to convert the token input after word splitting into a vector input, and here the approach is to use the shaping index corresponding to each token to obtain the corresponding column of vectors in the word encoding matrix. The $W_e\in\mathbb{R}^{d \times V}$ and $W_p\in\mathbb{R}^{d \times S}$ are the word encoding matrix and position encoding matrix respectively, $V$ is the size of the vocabulary list, $S$ is the maximum length of the input that can be accepted by the model, and $d$ is the size of the representation vector dimension. The vector encoding for each token is the sum of word encoding and position encoding. After word encoding, we obtain $\{z_{01}, z_{02}, ... , z_{0n}\}$ where $z_{0i}\in\mathbb{R}^{d}$. Next these word vectors will be passed through multiple Transformer Blocks, and it should be noted that the inputs and outputs of the Transformer Blocks passing through any of the layers will be of the exact same form, so that the final representation of each token after capturing the contextual information is obtained $\{z_{l1}, z_{l2}, ..., z_{ln}\}$,where$z_{li}\in\mathbb{R}^{d}$.

\section{Verify that the protein representation}\label{9}
\begin{figure}[htbp]
\centering
\includegraphics[scale=0.35]{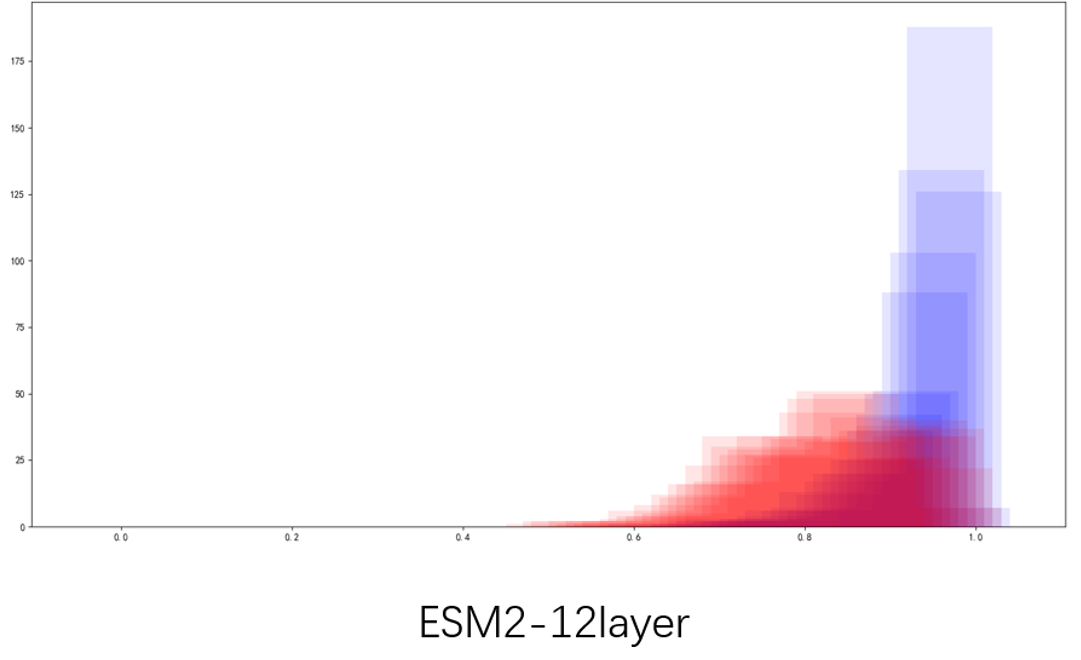}
\caption{ESM2 model cosine similarity distribution (positive samples in blue, negative samples in red)}
\label{homo1}
\end{figure}

\begin{figure}[htbp]
\centering
\includegraphics[scale=0.35]{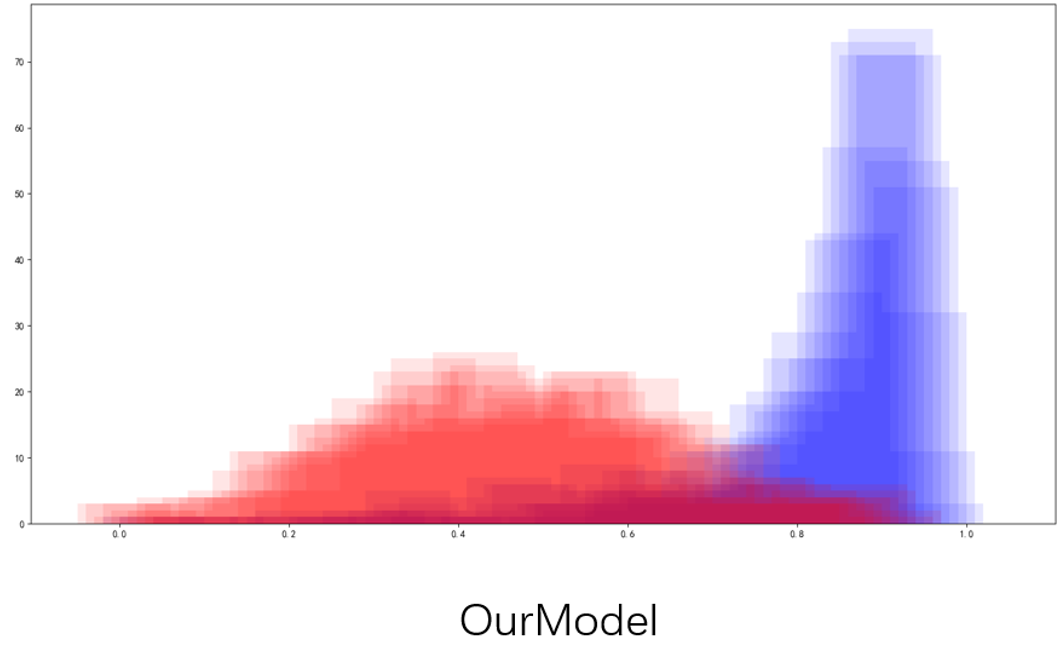}
\caption{ComproESM cosine similarity distribution (positive samples in blue, negative samples in red)}
\label{homo2}
\end{figure}

In order to further validate the changes produced by the protein representation, in this paper, 1010 pairs of proteins belonging to a family were randomly sampled as positive samples and an equal number of pairs of proteins that do not belong to a protein family were randomly sampled as negative samples, and the similarity between the pairs of proteins was calculated using cosine similarity. The similarity takes values between 0 and 1. The results of the two models are plotted in Figure\ref{homo1} and Figure\ref{homo2}. 

From the figure, it can be seen that the model proposed in this paper can clearly distinguish between positive and negative samples, and there is still a large part of overlapping space in the distribution of positive and negative samples for ESM2. Therefore, the biggest advantage of the proposed model is that it incorporates family information into the protein representation with little or no loss of the quality of the ESM2 representation, which itself carries the structural and functional information of the protein. Therefore, the protein representations generated using the model proposed in this paper can outperform the original ESM2-generated protein representations for a variety of protein downstream prediction tasks.

\section{Protein Classification}\label{1}
Search engines compute target sequences by sequence comparison algorithms to compare them with known sequences in databases to achieve classification, such as BLAST, and the performance of this method is limited by the comparison algorithm. Therefore, the use of machine learning methods to classify protein families based on protein sequences has become a hot research topic in recent years.
\begin{figure}[htbp]
\centering
\includegraphics[scale=0.5]{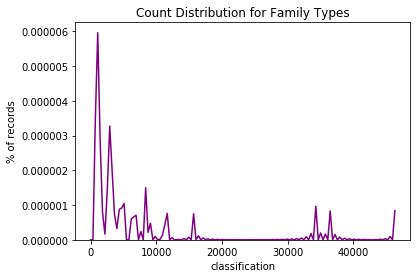}
\caption{Distribution of protein number in the protein classification }
\label{task_class_pic}
\end{figure}
\begin{table}[htbp]
\caption{Statistics on the number of proteins in each category in the Protein Classification Dataset}
\begin{center}
\small
\begin{tabular}{lcc}
\hline
Family & Number \\
\hline
HYDROLASE & 29,726\\
TRANSFERASE & 24,338\\
OXIDOREDUCTASE & 22,578\\
IMMUNE SYSTEM & 11,087\\
HYDROLASE/HYDROLASE INHIBITOR & 9569\\
LYASE & 8445\\
... & ...\\
ENDOCYTOSIS, PROTEIN BINDING & 1\\
Transferase, Cell Cycle & 1\\
HORMONE/GENE REGULATION & 1\\
\hline
\end{tabular}
\end{center}
\label{task_class_tab}
\end{table}

In this paper, we obtain the protein classification dataset on Kaggle, and organise the dataset, each sample is composed of protein sequences and their corresponding family names, and count the number of proteins corresponding to each family type (see table \ref{task_class_tab} and figure \ref{task_class_pic}), and it can be seen from figure \ref{task_class_pic}, it can be seen that the number of protein family distributions in this dataset shows a long-tailed distribution, so the top 250 classes of families with the highest number of proteins are taken as the predicted labels, and the final data obtained is 278,866. Considering the time of the test, this paper uniformly samples each class to one percent of the original one, and divides the training set and the test set according to the ratio of 1:1. Finally, 1394 training samples and test samples were obtained, and the model uses the simplest single-layer fully connected neural network to classify the input protein representations and use the classification accuracy for representation quality assessment. The test results are shown in Table \ref{metrics}.

\begin{figure}[htbp]
\centering
\includegraphics[scale=0.25]{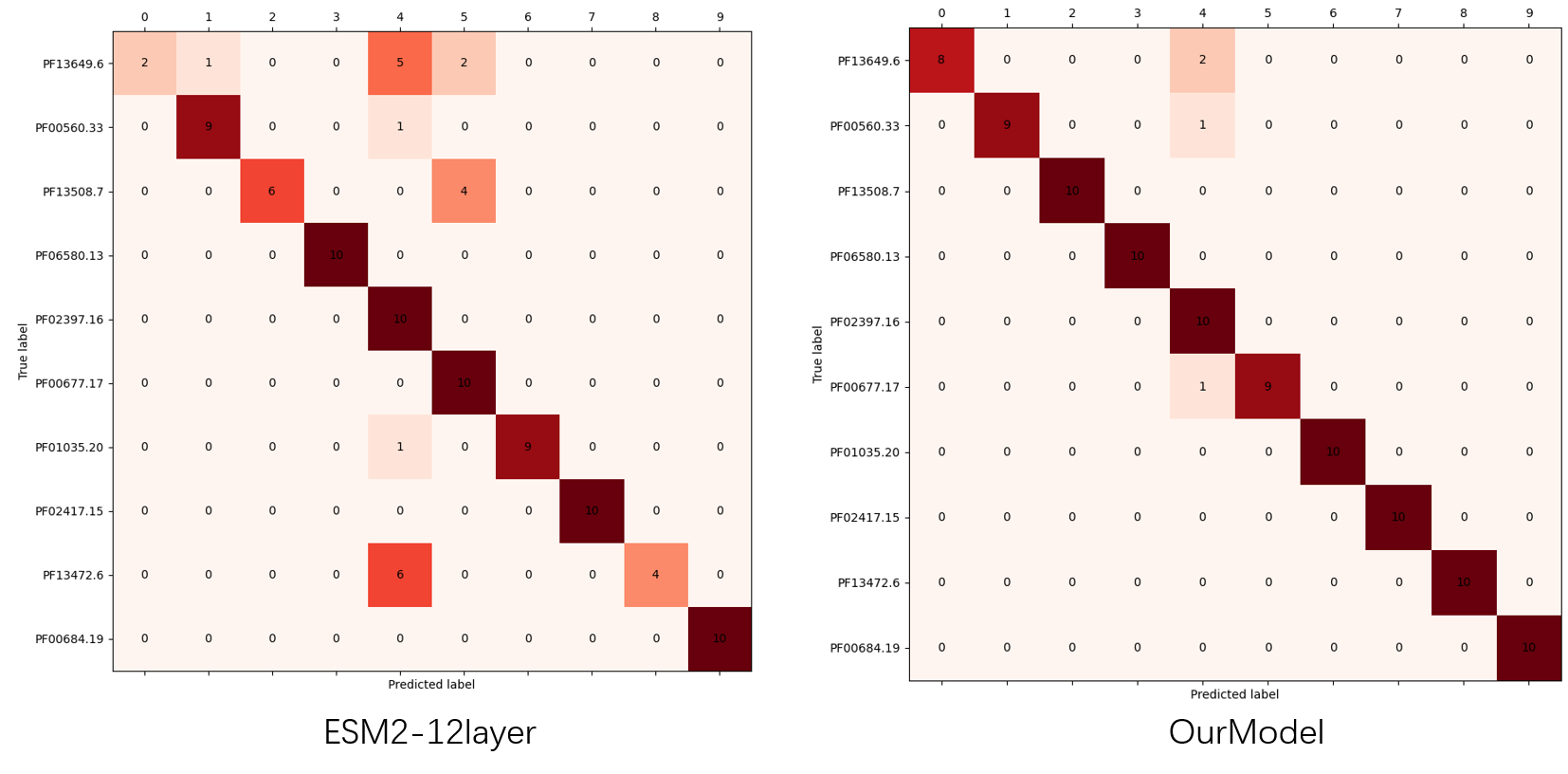}
\caption{Confusion matrix for protein classification tasks}
\label{confusion_pic}
\end{figure}

\section{Mutation Effects}\label{2}
Masked Language Modelling Objective The training objective of a trained protein language model is to output the probability of an amino acid occurring at a position in a protein given its surroundings. In this paper, we use this property to score mutations in amino acid sequences. For a given mutation, the amino acids in the original protein can be considered as a reference state, and the probability assigned to the mutated amino acid is compared to the probability assigned to the original amino acid thereby deriving the mutation score.

Specifically, the mutation score is evaluated using a log odds ratio at the mutation site, and when multiple mutation sites are present in a protein sequence, the scores at each position are summed using an addition method:
\begin{equation}
\sum_{t\in T}{\rm log}p(x_t=x_t^{mt}|x_{/T})-{\rm log}p(x_t=x_t^{wt}|x_{/T}),
\end{equation}
Here the sum is summed over all mutation positions, and the sequence input to the model is masked at each mutation position.

Riesselman et al.~\cite{riesselman2018deep} collected a set of 41 deep mutation scans and thus evaluated the model, these scans consisted of assessing the activity of a set of different proteins on various tasks. The experiments corresponded to different measurements in different tasks with different functions being tested. Each deep mutation scan dataset was treated as a separate prediction task and the model was used to score each variant in the dataset. These tasks were divided into a validation set consisting of ten mutation scan datasets and a test set consisting of the remaining datasets. Protein representation quality was assessed by comparing scores to experimental measurements using Spearman rank correlation.

Rank correlation, also known as rank correlation, is a non-parametric statistical method, but does not require a distribution of the original variable. It is suitable for information that does not follow a normal distribution, as well as information where the overall distribution is unknown and the original data is expressed in terms of rank.The Spearman rank correlation coefficient is used to calculate the correlation between two ordered sets of data. The formula for calculation is:
\begin{equation}
r_s = 1-\frac{6\sum{d^2_i}}{n(n^2-1)},
\end{equation}
where $d_i=(x_i-y_i)$, $x_i$ and $y_i$ denote the position of the two variables after sorting, and $n$ is the total number of samples.

The absolute value of Spearman's rank correlation coefficient is less than 1. The magnitude of the absolute value represents the strength of the correlation, and an absolute value of 1 indicates that the two series are perfectly correlated. When $r_s$ is positive, then the two sequences are positively correlated; when $r_s$ is negative, then the two sequences are negatively correlated.

The correlation between the mutation scores predicted by the model and the mutation scores generated by the real experiment is calculated, and the Pearman rank correlation is used here to return the correction. Correction is closer to 1, which means that the correlation between the two sequences is greater, and in this paper, we use the model to calculate the confusion degree of the amino acid sequences, so as to realise the zero-shot prediction of the activity of the protein. In this paper, we use the model to calculate the perplexity of amino acid sequence to achieve zero-shot prediction of protein activity, and the results are shown in Table \ref{metrics}.

\section{Activity Prediction}\label{3}
Protein mutations affect the activity of protein function, and protein activity was predicted using linear regression modelling of coded proteins regression, with the mean square error used to assess the prediction. The biological activity of protein mutations is predicted by training a simple mutation prediction network, i.e., using an embedded representation of the protein coding model to the amino acid sequence. In this paper, a simple multilayer perceptron is used to predict the mutation effect of protein sequences. The embedding representations of the dataset are saved by pre-computation. Afterwards a test of the downstream task is completed using these embedding representations and the activity corresponding to each representation.
In this test, the activity of ß-lactamase variants is predicted by training a model. The data was initially obtained from deep mutation scans and published by the Envision paper ~\cite{gray2018quantitative}. The dataset was taken from a fasta , and each entry in the file contains: the mutated ß-lactamase sequence, a residue in which the mutation occurs (exchanged for another amino acid) and a target value corresponding to the activity of the mutated protein, which describes the magnitude of the strength of the protein to act on a function and is a real-valued value. The quality of the protein representation was assessed by comparing the mean square error between the predicted activity values of the model and the experimentally measured activity values, and the results are shown in Table \ref{metrics}.

\section{Protein-protein interactions}\label{4}
PPIs have so far been studied from different perspectives (e.g. signalling, biochemistry, etc.) using a variety of approaches. Protein-protein interaction networks can then be constructed using this information. The use of machine learning methods to complete the complementation of large PPI networks and to make inference predictions on PPI networks is a major research topic in computational biology. In order to explore whether the protein representation model trained in this paper can embed protein interaction relationships into protein representation, the human protein-protein interaction network~\cite{agrawal2018large} provided by Stanford University is selected in this paper. This is a protein-protein interaction network that contains physical interactions between proteins in the experimentally recorded human body, such as metabolic enzyme coupling interactions and signalling interactions. The nodes on the graph represent human proteins and the edges represent physical interactions between proteins in human cells (see Table \ref{ppi_tab}).

Consider each edge in the table \ref{ppi_tab} as a positive example and use the randomly generated edges as negative examples. The final positive and negative examples are divided in the training and test sets in the ratio of 9 : 1, respectively. The two proteins were encoded into vectors with the model separately, spliced and then directly predicted by a binary neural network, with prediction accuracy as the evaluation metric, and the results are shown in Table \ref{metrics}.
\begin{table}[htbp]
\caption{Statistics of human protein-protein interaction}
\centering
\small
\begin{tabular}{lcc}
\hline
Attribute & Quantity \\
\hline
Nodes & 21,577 \\
Edges & 342,353 \\
SCC Node Count & 21,521 \\
SCC Edge Count & 342,316 \\
Triangle Count & 55,614,585 \\
Closed Triangle Ratio & 0.045652 \\
Diameter & 8 \\
\hline
\end{tabular}
\label{ppi_tab}
\end{table}

\section{Function Prediction}\label{5}
TheGeneOntology resource (\href{http://geneontology.org}{GO}) is the most comprehensive and widely used knowledge base on gene function. Therefore, in this paper, we test whether the protein embeddings generated by the protein representation model proposed in this paper can accurately predict the functions of protein sequences by collecting the functional descriptions of proteins from the Gene Ontology database. Functional descriptions corresponding to each class of proteins can be extracted from the protein database, and a total of 39,168 human proteins with 23,666 protein functions contained in all annotations were extracted from the UniprotKB database, and the model is required to accurately predict the correct functional option for a defined protein, given its functional description and three randomly generated functional descriptions. Detailed statistical information about the dataset is given in Table \ref{pro_func_tab}.

In this paper, we use PubMedBERT~\cite{gu2021domain} to obtain a representation of the functional description.PubMedBert is a pre-trained model trained from scratch from a large mixed dataset containing knowledge of a range of biological literature on diseases, drugs, genes, organs, cells, etc. PubMedBert uses masks on all of this data to predict the task. These data were pre-trained using the mask prediction task and fine-tuned and achieved state-of-the-art performance on a variety of tasks such as named entity recognition, PICO, relation extraction, semantic similarity, document classification, knowledge quizzing, and more. Each character-level encoding is obtained by feeding the functional description of the protein into PubMedBert, and all these encodings are averaged to obtain the encoding of the sentence. The sentence representation should contain rich semantic information, and a linear layer is used to project the protein representation into the representation space of the functional description. Afterwards, the similarity between the functional description and the protein representation is computed using vector dot product and the one with the highest similarity is selected as the correct option. The accuracy of selecting the correct option using the model is used as an evaluation metric of how good the protein representation is, and the results are shown in Table \ref{metrics}.
\begin{table}[htbp]
\caption{Protein function prediction dataset statistics}
\centering
\small
\begin{tabular}{lcc}
\hline
Attribute & Quantity \\
\hline
Protein Quantity & 39,168 \\
Protein function quantity & 23,666 \\
Number of Annotation Functions & 45,877 \\
Average Length of Functional Description & 69.327 \\
Functional Description Vocabulary & 30,522 \\
\hline
\end{tabular}
\label{pro_func_tab}
\end{table}

\section{Homology Detection}\label{6}
\begin{table*}[htbp]
\caption{Remote homology detection}
\centering
\small
\begin{tabular}{lcccccc}
\hline
Model & SF-Hit@10 & SF-Hit@5 & SF-Hit@1 & F-Hit@10 & F-Hit@5 & F-Hit@1 \\
\hline
ESM2-12layer & 47.04\% & 44.33\% & 28.52\% & 14.67\% & 9.53\% & 2.27\% \\
OurModel & 50.00\% & 47.43\% & 31.85\% & 17.47\% & 10.14\% & 2.72\% \\
\hline
\end{tabular}
\label{remote}
\end{table*}
We use SCOPe~\cite{fox2014scope} to assess remote homology detection. According to standard practice~\cite{soding2011protein}, Rossman-type folds (c.2-2.5, c.27 and 28, c.30 and 31) and four- to eight-bladed propellers (b.66-b.70) were excluded from this paper. Proteins from the same superfamily belong to different families. Vector inner products were used directly to measure the structural and functional similarity of the protein representations, taking the highest 10 scored proteins to consider them as remotely homologous to the target proteins, using the Hit-10 metric, i.e., the proportion of proteins corresponding to the top 10 scores and the target proteins that do belong to the same protein superfamily. The experiments are performed by measuring the density of homologous proteins near the query sequence based on an unsupervised classifier that measures the similarity between protein representations based on the inner product of vectors. For each domain, a vector similarity query will be performed on all other domains and they will be sorted by distance to the query domain. For superfamily level assessment, any domains with the same superfamily and not belonging to the same family are considered as positive examples; any domains with different superfamilies are considered as negative samples. The metrics were measured using Hit-10, which gives the proportion of correctly predicted remote homologues among the ten highest ranked results. The experimental data is consistent with the ESM2 paper, and the final results are shown in Table. \ref{metrics}
In order to distinguish the effect of the model more clearly, this paper also increased the difficulty of the experiments by using only protein sequences with sequence similarity below 40\% to participate in the experiments, and at the same time, due to the limitation of computational resources, only proteins of the $\alpha$ folding family were used. The experiment verifies that the protein representation can detect remote homology between proteins belonging to the same superfamily (but belonging to different families) as well as between proteins belonging to the same fold (but belonging to different superfamilies). The final sample size of the experiment was 2644, and the correlation between two two proteins was calculated using cosine similarity, and the top 10 proteins were taken as the homologous proteins detected by the model. The final calculations were done separately (Hit@10: the proportion of proteins ranked in the top 10 of similarity that are homologous proteins, Hit@5 and Hit@1), and the final results are shown in Table \ref{remote}, where SF stands for the results on superfamily homology detection, and F stands for the results on folding homology detection.

From Table \ref{remote}, it can be seen that when the sequence similarity of the proteins in the dataset used for testing is reduced to below 40\%, the accuracy of the model on the remote homology detection task will be greatly reduced, and the model proposed in this paper performs significantly better than the ESM2 on this task as well.It is argued that this result arises due to the fact that the model proposed in this paper is introduced into the training of the protein family data, the protein representation itself spatially conforms to the reticulation distribution of protein families, in other words, the protein representation contains the functional information of proteins, so that functionally similar proteins can still be recognised better when sequence similarity decreases. Both models perform poorly in protein folding homology detection when sequence similarity decreases, probably because the folding structure of proteins itself is highly correlated with sequence similarity.

\end{document}